\documentclass[12pt,preprint]{aastex}

\slugcomment{To appear in ApJ (February 2006)}

\shorttitle{Fundamentals of Speckle Nulling}
\shortauthors{Bord\'e \& Traub}

\begin{document}

\title{High-contrast Imaging from Space: \\
       Speckle Nulling in a Low Aberration Regime}

\author{Pascal J. Bord\'e\altaffilmark{1}}
\affil{Harvard-Smithsonian Center for Astrophysics, \\
       60 Garden Street, Cambridge, MA 02138}
\email{pborde@cfa.harvard.edu}
\altaffiltext{1}{Michelson Postdoctoral Fellow; present address: Michelson Science Center, California Institute of Technology, 770 S Wilson Avenue, Pasadena, CA 91125}

\author{Wesley A. Traub\altaffilmark{2}}
\affil{Jet Propulsion Laboratory, M/S 301-451, \\
       4800 Oak Grove Drive, Pasadena, CA 91109}
\email{wtraub@jpl.nasa.gov}
\altaffiltext{2}{Research Associate, Harvard-Smithsonian Center for Astrophysics}

\begin{abstract}
High-contrast imaging from space must overcome two major noise sources to successfully detect a terrestrial planet angularly close to its parent star: photon noise from diffracted star light, and speckle noise from star light scattered by instrumentally-generated wavefront perturbation. Coronagraphs tackle only the photon noise contribution by reducing diffracted star light at the location of a planet. Speckle noise should be addressed with adaptative-optics systems. Following the tracks of \citet{Malbet95}, we develop in this paper two analytical methods for wavefront sensing and control that aims at creating \emph{dark holes}, i.e. areas of the image plane cleared out of speckles, assuming an ideal coronagraph and small aberrations. The first method, \emph{speckle field nulling}, is a fast FFT-based algorithm that requires the deformable-mirror influence functions to have identical shapes. The second method, \emph{speckle energy minimization}, is more general and provides the optimal deformable mirror shape via matrix inversion. With a $N\!\times\!N$ deformable mirror, the size of matrix to be inverted is either $N^2\!\times\!N^2$ in the general case, or only $N\!\times\!N$ if influence functions can be written as the tensor product of two one-dimensional functions. Moreover, speckle energy minimization makes it possible to trade off some of the dark hole area against an improved contrast. For both methods, complex wavefront aberrations (amplitude and phase) are measured using just three images taken with the science camera (no dedicated wavefront sensing channel is used), therefore there are no non-common path errors. We assess the theoretical performance of both methods with numerical simulations including realistic speckle noise and experimental influence functions. We find that these speckle nulling techniques should be able to improve the contrast by several orders of magnitude.
\end{abstract}

\keywords{Instrumentation: adaptive optics --- techniques: high angular resolution --- planetary systems}

%
%
\section{Introduction} \label{sec:introduction}
The field of extrasolar planet research has recently made a leap forward with the direct detection of extrasolar giant planets (EGPs). Using Spitzer Space Telescope, \citet{Charbonneau05} and \citet{Deming05} have detected infrared photons from two transiting planets, TrES-1b and HD209458b, respectively. \citet{Chauvin04,Chauvin05} have reported the infrared imaging of an EGP orbiting the nearby young brown dwarf 2M1207 with VLT/NACO, whereas \citet{Neuhauser05} have collected evidence for an EGP companion to the T-Tauri star GQ Lup using VLT/NACO as well. 

Although there are claims that the direct detection of terrestrial planets could be performed from the ground with -- yet to come -- extremely large telescopes \citep{Angel03,Chelli05}, it is widely believed that success will be more likely in space. Direct detection is the key to spectroscopy of planetary atmospheres and discovery of biomarkers, namely indirect evidence of life developed at the planetary scale \citep[e.g.][]{DesMarais02}. 

Both NASA and ESA have space mission studies well underway to achieve this task. Darwin, the European mission to be launched in 2015, will be a thermal infrared nulling interferometer with three 3.5-m free-flying telescopes \citep{Karlsson04}. Terrestrial Planet Finder, the American counterpart, will feature two missions: a $8\!\times\!3.5$~m-monolithic visible telescope equipped with a coronagraph (TPF-C) to be launched in 2015, and an analog to Darwin (TPF-I) to be launched in the 2015--2019 range \citep{Coulter04}.

The direct detection of the photons emitted by a terrestrial planet is made very challenging by the angular proximity of the parent star, and by the very high contrast (i.e. luminosity ratio) between the planet and its star: about $10^6$ in the thermal infrared and about $ 10^{10}$ in the visible. Both wavelength ranges have their scientific merits and technical difficulties, and both of them are thought to be necessary for an unambiguous detection of habitability and signs of life \citep[e.g.][]{DesMarais02}. In this paper, we deal with the visible range only.

In the visible, planet detection faces two fundamental noise sources: (i) quantum noise of the diffracted star light, and (ii) speckle noise due to the scattering of the star light by optical defects. \citet{Labeyrie95} proposed a technique based on \emph{dark speckles} to overcome speckle noise: random fluctuations of the atmosphere cause the speckles to interfere destructively and disappear at certain locations in the image, thus creating localized dark spots suitable for planet detection. The statistical analysis of a large number of images then reveals the planet as a spot persistently brighter than the background.

\citet{Malbet95} proposed to use a deformable mirror (DM) instead of the atmosphere to make speckles interfere destructively in a targeted region of the image called \emph{search area} or \emph{dark hole} (DH or $\mathcal{H}$). Following the tracks of these authors, this paper discusses methods to reduce the speckle noise below the planet level by using a DM and an ideal coronagraph. However, unlike \citet{Malbet95}, we propose non-iterative algorithms, in order to limit the number of long exposures needed for terrestrial planet detection. We will refer to these methods as \emph{speckle nulling} techniques, as \cite{Trauger04} call them. Technical aspects of this work are inspired by the High Contrast Imaging Testbed \cite[HCIT;][]{Trauger04}, a speckle-nulling experiment hosted at the Jet Propulsion Laboratory, specifically designed to test TPF-C related technology.

After reviewing the process of speckle formation to establish our notations (\S\ref{sec:speckle_formation}), we derive two speckle nulling methods in the case of small aberrations (\S\ref{sec:speckle_nulling}). The speckle nulling phase is preceded by the measurement of the electric field in the image plane (\S\ref{sub:measurement}). The performance of both methods are then evaluated with one- and two-dimensional simulations (\S\ref{sec:simulations}), first with white speckle noise (\S\ref{sub:sim_white}), then with non-white speckle noise (\S\ref{sub:sim_real}). Various effects and instrumental noises are considered in \S\ref{sec:discussion}. Finally, we conclude and discuss some future work (\S\ref{sec:conclusion}).

%
%
\section{Speckle formation} \label{sec:speckle_formation}
This paper is written in the framework of Fourier optics considering a single wavelength, knowing that a more sophisticated theory (scalar or vectorial) in polychromatic light will eventually be needed. Fourier transforms (FTs) are signaled by a hat.

Let us consider a simple telescope with an entrance pupil $\mathcal{P}$. In the pupil plane, we use the reduced coordinates $(u,v) = (x/\lambda,y/\lambda)$, where $(x,y)$ are distances in meters and $\lambda$ is the wavelength. We define the pupil function by
\begin{equation} \label{eq:P}
P(u,v) \equiv
\left \{
\begin{array}{l}
1 \mbox{ if } (u,v) \in \mathcal{P}, \\
0 \mbox{ otherwise.}
\end{array}
\right.
\end{equation}

Even in space, i.e. when not observing through a turbulent medium like the atmosphere, the optical train of the telescope is affected by phase and amplitude aberrations. Phase aberrations are wavefront corrugations that typically originate in mirror roughness caused by imperfect polishing, while amplitude aberrations are typically the result of a heterogeneous transmission or reflectivity. Moreover, Fresnel propagation turns phase aberrations into amplitude aberrations, and the reverse \citep[e.g.][]{Guyon05b}. Regardless of where they originate physically, all phase and amplitude aberrations can be represented by a complex aberration function $\phi$ in a re-imaged pupil plane, so that the aberrated pupil function is now $P e^{i \phi}$.

The electric field associated with an incident plane wave of amplitude unity is then
\begin{equation} \label{eq:E_pu}
E(u,v) = P(u,v)\,e^{i \phi(u,v)}.
\end{equation}

Exoplanet detection requires that we work in a regime where aberrations are reduced to a small fraction of the wavelength. Once in this regime, we can replace $e^{i \phi}$ by its first order expansion  $1 + i \phi$ (we will discuss in \S\ref{sub:discuss_linearity} the validity of this approximation). The electric field in the image plane being the FT of (\ref{eq:E_pu}), we get
\begin{equation} \label{eq:E_im}
\widehat{E}(\alpha,\beta) = \widehat{P}(\alpha,\beta) + i \, \widehat{P\phi}(\alpha,\beta),
\end{equation}
where $(\alpha,\beta)$ are angular coordinates in the image plane.

The physical picture is as follows. The first term ($\widehat{P}$) is the direct image of the star. The second term ($\widehat{P\phi}$) is the field of speckles surrounding the central star image, where each speckle is generated by the equivalent of first-order scattering from one of the sinusoidal components of the complex aberration $\phi$. Each speckle is essentially a ghost of the central PSF.

In the remainder of this paper, we focus on means to measure and correct the speckles in a coronagraphic image. Following \cite{Malbet95} we will leave out the unaberrated PSF term by assuming that it is was canceled out by a coronagraph of some sort (see \citet{Quirrenbach05} for a review on coronagraphs). Thus we clearly separate the gain in contrast that can be obtained by reducing the diffracted light with the coronagraph on one hand, and by fighting the scattered light with the speckle nulling technique on the other hand.

%
%
%
\section{Speckle nulling theory} \label{sec:speckle_nulling}
The purpose of speckle nulling is to reduce the speckle noise in a central region of the image plane. This region, the dark hole, then becomes dark enough to enable the detection of companions much fainter than the original speckles. Speckle nulling is achieved by way of a servo system that has a deformable mirror as actuator. Because our sensing method requires DM actuation and is better understood with the knowledge of the command control theory, we first model the deformable mirror (\S\ref{sub:deformable_mirror}), then present two algorithms for the command control (\S\ref{sub:field_nulling} \& \ref{sub:energy_min}), and conclude with the sensing method (\S\ref{sub:measurement}).

%
%
\subsection{Deformable mirror} \label{sub:deformable_mirror}
The deformable mirror (DM) in \cite{Trauger03} consists of a continuous facesheet supported by $N\!\times\!N$ actuators arranged in a square pattern of constant spacing. This DM format is well adapted to either square or circular pupils, the only pupil shapes that we consider in this paper\footnote{Two square DMs can be assembled to accommodate an elliptical pupil such as the one envisioned for TPF-C.}. We assume that the DM is physically located in a plane that is conjugate to the entrance pupil. However, what we call DM in the following is the projection of this real DM in the entrance pupil plane. The projected spacing between actuators is denoted by $d$. We assume that the optical magnification is such that the DM projected size is matched to the entrance pupil, i.e. $Nd = D$, where $D$ is either the pupil side length or its diameter. The DM surface deformation in response to the actuation of actuator ${(k,l)  \in \{0 \ldots N\!-\!1 \}^2}$ is described by an \emph{influence function}, denoted by $f_{kl}$. The total phase change introduced by the DM (DM phase function) is
\begin{equation} \label{eq:psi}
\psi(u,v) \equiv \sum_{k=0}^{N-1} \sum_{l=0}^{N-1} a_{kl}\,f_{kl}(u,v),
\end{equation}
where $a_{kl}$ are actuator strokes (measured in radians). Note that contrary to the complex aberration function $\phi$, the DM phase function is purely real.

With an ideal coronagraph and a DM, the image-plane electric field formerly given by (\ref{eq:E_im}) becomes
\begin{equation} \label{eq:E_im2}
\widehat{E}'(\alpha,\beta) = i\,\widehat{P\phi}(\alpha,\beta) + i\,\widehat{P\psi}(\alpha,\beta).
\end{equation}

In the next two sections, we explore two approaches for speckle nulling. In \S\ref{sub:field_nulling}, we begin naively by trying to cancel $\widehat{E}'$. Because there is a maximum spatial frequency that the DM can correct for, the DH has necessarily a limited extension. Any energy at higher spatial frequencies will be aliased in the DH and limit its depth. Therefore, the DM cannot be driven to cancel $\widehat{E}'$, unless $\widehat{P\phi}$ is equal to zero outside the DH (i.e. unless there are already no speckles outside the DH). With this in mind, we start over in \S\ref{sub:energy_min} with the idea that speckle nulling is better approached by minimizing the field energy.

%
%
\subsection{Speckle field nulling} \label{sub:field_nulling}
The speckle field nulling approach consists in trying to null out $\widehat{E}'$ in the DH region ($\mathcal{H}$), meaning we seek a solution to the equation
\begin{equation} \label{eq:field1}
\forall (\alpha,\beta) \in \mathcal{H}, \quad \widehat{P\phi}(\alpha,\beta) + \widehat{P\psi}(\alpha,\beta) = 0,
\end{equation}
although, as we shall show, this equation has no exact solution unless $\widehat{P\phi}$ happens to be a band-limited function within the controllable band of the DM.

By replacing $\psi$ with its expression (\ref{eq:psi}), we obtain
\begin{equation} \label{eq:field2}
\forall (\alpha,\beta) \in \mathcal{H}, \quad \sum_{k=0}^{N-1} \sum_{l=0}^{N-1} a_{kl}\,\widehat{Pf_{kl}}(\alpha,\beta)
= - \widehat{P\phi}(\alpha,\beta).
\end{equation}
We recognize in (\ref{eq:field2}) a linear system in the $a_{kl}$ that could be solved using various techniques such as singular value decomposition \citep[SVD;][\S2.6]{Press02}. Although general, this solution does not provide much insight in the problem of speckle nulling. For this reason, let us examine now a different solution, less general but with more explanatory power. We will comment on the use of SVD at the end of this section.

We consider a square pupil. In this case, all DM actuators receive light and the pupil function has no limiting effect on the DM phase function, i.e. $P\psi = \psi$. Moreover, we assume all influence functions to be identical in shape, and write $f_{kl}(u,v) = f(u-k\frac{d}{\lambda},v-l\frac{d}{\lambda})$. Under these hypotheses, 
\begin{equation} \label{eq:psi2}
P\psi(u,v) = f(u,v) \ast \sum_{k=0}^{N-1} \sum_{l=0}^{N-1} a_{kl} \, \delta \! \left( u - k \frac{d}{\lambda}, v - l \frac{d}{\lambda} \right),
\end{equation}
where $\delta$ is Dirac's bidimensional distribution, and $\ast$ denotes the convolution.

Substituting $\widehat{P\psi}$ by the FT of (\ref{eq:psi2}) in (\ref{eq:field1}) yields
\begin{equation} \label{eq:field3}
\forall (\alpha,\beta) \in \mathcal{H}, \quad \sum_{k=0}^{N-1} \sum_{l=0}^{N-1} a_{kl}\,e^{-i \frac{2\pi d}{\lambda} (k \alpha + l \beta)}
= - \frac{\widehat{P\phi}(\alpha,\beta)}{\hat{f}(\alpha,\beta)}.
\end{equation}
We recognize in the left-hand side of (\ref{eq:field3}) a truncated Fourier series. If we choose the $a_{kl}$ to be the first $N^2$  Fourier coefficients of $-\widehat{P\phi}/\hat{f}$, i.e.
\begin{equation} \label{eq:coef1}
a_{kl} = \frac{2d^2}{\lambda^2} \int\!\!\!\int_{{[-\frac{\lambda}{2d}, \frac{\lambda}{2d}]}^2} -\frac{\widehat{P\phi}(\alpha,\beta)}{\hat{f}(\alpha,\beta)} \:
e^{i \frac{2\pi d}{\lambda} (k \alpha + l \beta)} \, \mathrm{d}\alpha \, \mathrm{d}\beta,
\end{equation}
then according to Fourier theory, we minimize the mean-square error between both sides of the equation \cite[see e.g.][\S1.5]{Hsu67}. This error cannot be reduced to zero unless the Fourier coefficients of $-\widehat{P\phi}/\hat{f}$ happen to vanish for $k,l < 0$ and $k,l > N$. At this point, we have reached the important conclusion that \emph{perfect speckle cancellation cannot be achieved with a finite-size DM unless the wavefront aberrations are band-limited}. Moreover, we can assert that the maximum DH extension is the square domain $\mathcal{H} \equiv {[-\frac{\lambda}{2d}, \frac{\lambda}{2d}]}^2 = {[-\frac{N}{2}\frac{\lambda}{D}, \frac{N}{2}\frac{\lambda}{D}]}^2$.

Solution (\ref{eq:coef1}) is physically acceptable only if the Fourier coefficients are real numbers, which means mathematically that $\widehat{P\phi}/\hat{f}$ should be Hermitian\footnote{A function $f$ is said to be Hermitian if $ \forall (x,y), \: f(x,y) = f^\ast(-x,-y)$. The FT of a real function is Hermitian and vice versa.}. If there are phase aberrations only, $P\phi$ is real, $\widehat{P\phi}/\hat{f}$ is Hermitian, and the $a_{kl}$ are real. This is no longer true if there are amplitude aberrations as well, reflecting the fact that the DM alone cannot correct both phase and amplitude aberrations in $\mathcal{H}$. However, by considering the Hermitian function that is equal to $\widehat{P\phi}/\hat{f}$ in one half of the DH, say $\mathcal{H}^+ \equiv [0,\frac{\lambda}{2d}] \times [-\frac{\lambda}{2d},\frac{\lambda}{2d}]$, we obtain the real coefficients
\begin{equation} \label{eq:coef2}
a_{kl} = 4d^2 \int\!\!\!\int_{\mathcal{H}^+} -\frac{\widehat{P\phi}(\alpha,\beta)}{\hat{f}(\alpha,\beta)} \:
\cos \! \left[ \frac{2\pi d}{\lambda} (k \alpha + l \beta) \right] \mathrm{d}\alpha \, \mathrm{d}\beta,
\end{equation}
that correct both amplitude and phase aberrations in $\mathcal{H}^+$. As we have $\frac{\lambda}{2d} = \frac{N}{2}\frac{\lambda}{D}$, the DH has a size of $N\!\times\!N$ resolution elements (resels) with phase aberrations only, and of $\frac{N}{2}\!\times\! N$ resels with phase and amplitude aberrations. Therefore, \emph{a DM can correct both amplitude and phase aberrations in the image plane, albeit in a region that is either the left, right, upper, or lower half of the phase-corrected region.}

As \citet{Malbet95} pointed out, let us remind the reader that $\frac{\lambda}{2d}$ is equal to the Nyquist frequency for a sampling interval $\frac{d}{\lambda}$. Therefore, we find that maximum extension for the DH corresponds to the range where the sampling theorem applies to the wavefront at the DM actuator scale. Indeed, taking the inverse FT of (\ref{eq:field3}) leads to the wavefront reconstruction formula
\begin{equation} \label{eq:field4}
P\phi(u,v) = - \sum_{k=0}^{N-1} \sum_{l=0}^{N-1} a_{kl} \, f \! \left( u - k \frac{d}{\lambda}, v - l \frac{d}{\lambda} \right).
\end{equation}
Again, this reconstruction cannot be perfect unless the spectrum of $P\phi$ is contained in $\mathcal{H}$. Note that because $\hat{f}$ is generally not a flat function (as it would be the case if influence functions were for instance 2D sinc functions), actuator strokes are not equal to the negative of wavefront values sampled at the actuator locations.

Our Fourier solution was derived by assuming that (a) all influence functions are identical in shape, and (b) that the pupil has a square shape. Hypothesis (a) appears to be reasonable at least for the DM in use on the HCIT (Joseph Green, personal communication), but this remains to be precisely measured. If hypothesis (b) is relaxed then (i) some actuators do not receive any light and play no role, so there are effectively fewer terms in the summation in (\ref{eq:field3}), and (ii) the fact that influence functions on the pupil boundary are only partly illuminated is ignored.

Now that we have two methods to solve (\ref{eq:field1}), Fourier expansion and SVD, let us compare their solutions. We deal here with functions belonging to the Hilbert space of square integrable functions $f\!: \mathcal{H} \rightarrow \mathbb{C}$. This space has $<f,g> \equiv \int\!\!\!\int_\mathcal{H} f \, g^\ast$ for dot product, and $||f|| \equiv \sqrt{\int\!\!\!\int_\mathcal{H} |f|^2}$ for norm. As mentioned earlier, Fourier expansion minimizes the mean-square error between both sides of (\ref{eq:field3}), i.e. $||(\widehat{P\phi}+\widehat{P\psi})/\hat{f}||^2$. By contrast, SVD has the built-in property of minimizing the norm of the residuals of (\ref{eq:field2}), i.e. $||\widehat{P\phi}+\widehat{P\psi}||$. In other words, SVD minimizes $||\widehat{E'}||^2$, the speckle field energy, which seems more satisfactory from a physical point of view. To find out what is best, we have performed one-dimensional numerical simulations. It turns out that SVD yields dark holes 50\,\% deeper (median value) than Fourier expansion. In addition, SVD does not require all influence functions to have the same shape.

However, considering four detector pixels per resel in two dimensions (critical sampling), SVD would require us to manipulate matrices as large as $N^2\!\times\!4N^2$ (or even $N^2\!\times\!8N^2$ when real and imaginary parts are separated). Such matrices would occupy 537~MB of memory space for $64\!\times\!64$ actuators and single-precision floating-point numbers. By contrast, Fourier expansion would be straightforwardly obtained with FFTs of $2N\!\times\!2N$ arrays at critical sampling, but again at the cost of a 50\,\% shallower dark hole and a strong hypothesis on the influence functions.

In the next section, we seek to find a computationally less intensive solution that still minimizes the speckle energy in the dark hole, but does not require any hypothesis on the influence functions.

%
%
\subsection{Speckle energy minimization} \label{sub:energy_min}
Let us start with the idea that the best solution is defined as the one \emph{minimizing the total energy of the speckle field in the DH}. For the sake of simplicity, we assume once again a square pupil, but not necessarily a common shape for the influence functions. The total energy in the speckle field reads
\begin{equation} \label{eq:energy1}
\mathcal{E} \equiv \int\!\!\!\int_\mathcal{H} |\widehat{P\phi}(\alpha,\beta) + \widehat{\psi}(\alpha,\beta)|^2 \, \mathrm{d}\alpha \, \mathrm{d}\beta
\; = \; <\widehat{P\phi} + \widehat{\psi}, \widehat{P\phi} + \widehat{\psi}>,
\end{equation}
using the same notation as in \S\ref{sub:field_nulling}.

Given that $\partial \widehat{\psi}/\partial a_{kl} = \hat{f}_{kl}$, the energy is minimized when
\begin{equation} \label{eq:energy2}
\forall (k,l) \in {\{0 \ldots N\!-\!1\}}^2, \quad \frac{\partial \mathcal{E}}{\partial a_{kl}} = 0
\quad \Longleftrightarrow \quad
\Re \left( <\widehat{P\phi} + \widehat{\psi}, \hat{f}_{kl}> \right) = 0,
\end{equation}
where $\Re$ stands for the real part. Note that this is less demanding than (\ref{eq:field1}), as (\ref{eq:field1}) implies (\ref{eq:energy2}) but the reverse is not true.

Using the definition (\ref{eq:psi}) for $\psi$ and realizing that $<\hat{f}_{nm},\hat{f}_{kl}>$ is a real number\footnote{This property stems from the Hermitian character of $\hat{f}_{kl}$ together with the symmetry of $\mathcal{H}$.}, we get finally
\begin{equation} \label{eq:energy3}
\forall (k,l) \in {\{0 \ldots N\!-\!1\}}^2, \quad
\sum_{n=0}^{N-1} \sum_{m=0}^{N-1} a_{nm} <\hat{f}_{nm},\hat{f}_{kl}> \; = - \Re \left( <\widehat{P\phi},\hat{f}_{kl}> \right).
\end{equation}

As in (\ref{eq:field2}), we find a system that is linear in the actuator strokes. By replacing double indices with single ones, e.g. $(k,l)$ becomes $s = k \, N + l$, (\ref{eq:energy3}) can be solved in matrix format by inverting a $N^2\!\times\!N^2$ real matrix. This is already an improvement with respect to the $N^2\!\times\!4N^2$ complex matrix required by SVD in the previous section.

It appears that the same solution can be obtained with a much less demanding ${N\!\times\!N}$ matrix inversion, provided two-dimensional influence functions can be written as the tensor product of two one-dimensional functions (separation of variables), i.e. $f_{kl}(u,v) = g_k(u) \, g_l(v)$. This would be the case for box functions or bidimensional Gaussians, and is good at the 5\,\% level for the DM in use on the HCIT. This property also holds in the image plane since the FT of the previous equation yields $\hat{f}_{kl}(\alpha,\beta) = \hat{g}_k(\alpha) \, \hat{g}_l(\beta)$.

By separating variables, (\ref{eq:energy3}) becomes
\begin{equation} \label{eq:energy4}
\forall (k,l) \in {\{0 \ldots N\!-\!1\}}^2, \quad
\sum_{n=0}^{N-1} <\hat{g}_n,\hat{g}_k> \sum_{m=0}^{N-1} a_{nm} <\hat{g}_m,\hat{g}_l> \; = - \Re \left( <\widehat{P\phi},\hat{f}_{kl}> \right).
\end{equation}
As the left-hand side happens to be the product of three $N\!\times\!N$ matrices, (\ref{eq:energy4}) can be rewritten as an equality between square matrices.
\begin{equation} \label{eq:energy5}
G \, A \, G = \Phi,
\quad \mbox{where} \quad
\left \{
\begin{array}{l}
G_{kl} = \; <\hat{g}_k,\hat{g}_l> \\
A_{kl} = a_{kl} \\
\Phi_{kl} = - \Re \left( <\widehat{P\phi},\hat{f}_{kl}> \right).
\end{array}
\right.
\end{equation}
For square-box and actual HCIT influence functions, numerical calculations show that $G$ is diagonally dominant\footnote{A matrix $A=[a_{ij}]$ is said to be diagonally dominant if $\forall i, \: |a_{ii}| > \sum_{j \neq i} |a_{ij}|$.} and therefore invertible by regular Gaussian elimination. The solution to (\ref{eq:energy5}) is then
\begin{equation} \label{eq:energy6}
A = G^{-1} \, \Phi \, G^{-1}.
\end{equation}
Note that $G^{-1}$ can be precomputed and stored, so that computing the strokes effectively requires only two matrix multiplications. As shown in appendix~\ref{app:global}, an equivalent result can be obtained by working with pupil plane quantities.

As for the field nulling approach, correcting amplitude errors as well implies restricting the dark hole to either $\mathcal{H}^+ = [0,\frac{N}{2}\frac{\lambda}{D}] \times [-\frac{N}{2}\frac{\lambda}{D},\frac{N}{2}\frac{\lambda}{D}]$ or $\mathcal{H}^- = [-\frac{N}{2}\frac{\lambda}{D},0] \times [-\frac{N}{2}\frac{\lambda}{D},\frac{N}{2}\frac{\lambda}{D}]$. To account for amplitude errors and keep the formalism we have presented so far, it is sufficient to replace $\widehat{P\phi}$ by a function equal to $\widehat{P\phi}(\alpha,\beta)$ in either $\mathcal{H}^+$ or $\mathcal{H}^-$ (depending on the half where one wishes to create the dark hole), and equal to $\widehat{P\phi}^\ast(-\alpha,-\beta)$ in the other half (Hermitian symmetry). Because its FT is Hermitian, the new aberration function in the pupil plane is real, and thus the algorithm processes amplitude and phase errors at the same time as if there were phase errors only.

Let us derive the residual total energy in the DH after the correction has been applied. Starting from definition (\ref{eq:energy1}) and rewriting condition (\ref{eq:energy2}) as ${\Re ( <\widehat{P\phi} + \widehat{\psi}, \widehat{\psi}> ) = 0}$, we find
\begin{equation}
\mathcal{E}_\mathrm{min} = \; <\widehat{P\phi},\widehat{P\phi}> - <\widehat{\psi},\widehat{\psi}>.
\end{equation}
The former term is the initial speckle energy in the DH, while the latter is the speckle energy decrease gained with the DM. Mathematically, $\sqrt{\mathcal{E}_\mathrm{min}}$ measures the distance (according to the norm we have defined) between the speckle field and its approximation with the DM inside the DH. Because there is no exact solution to ($\ref{eq:field1}$), the residual energy cannot be made equal to zero in $\mathcal{H}^+$ or $\mathcal{H}^-$. However, the energy approach offers an additional degree of freedom: by reducing concentrically the domain over which the energy is minimized, the speckle energy can be further decreased (see \S\ref{sec:simulations}).

%
%
\subsection{Speckle field measurement} \label{sub:measurement}
So far, our speckle nulling theory has presupposed the knowledge of the speckle field $\widehat{P\phi}$, or equivalently of the phase and amplitude aberrations across the pupil, embodied in the complex phase function $P\phi$. In this section, we show how the speckle field can be measured directly in the image plane. As the detector measures an intensity, a single image yields only the modulus of the speckle field. The phase of the speckle field can be retrieved by perturbing the phase function $P\phi$ in a controlled way, and by recording the corresponding images, a process analogous to \emph{phase diversity} \citep[e.g.][]{Lofdahl94}. In our system, the DM provides the natural means for creating this controlled perturbation.

As we will see, exactly three images obtained with well chosen DM settings provide enough information to measure $\widehat{P\phi}$. Let us call image 0 the original image recorded with a setting $\psi_0$, whereas images 1 and 2 are recorded with settings $\psi_0+\delta\psi_1$ and $\psi_0+\delta\psi_2$.

To be general, we consider in the field of view the presence of an exoplanet and an exozodiacal cloud (exozodi for short), in addition to the star itself. The electric fields of these objects are incoherent with that of the star, so their intensities should be added to the star's intensity. Because they are much fainter than the star, the speckles they produce are negligible with respect to the star speckles, and their intensities can be considered as independent of $\phi$ and $\psi$. The total intensity of every image pixel $(\alpha,\beta)$ takes then the successive values 
\begin{equation} \label{eq:I_system1}
\left \{
\begin{array}{l}
I_0 = |\widehat{P\phi} + \widehat{\psi}_0|^2 + I_\mathrm{p} + I_\mathrm{z} \\
I_1 = |\widehat{P\phi} + \widehat{\psi}_0 + \widehat{\delta\psi_1}|^2 + I_\mathrm{p} + I_\mathrm{z} \\
I_2 = |\widehat{P\phi} + \widehat{\psi}_0 + \widehat{\delta\psi_2}|^2 + I_\mathrm{p} + I_\mathrm{z}, \\
\end{array}
\right.
\end{equation}
where $I_\mathrm{p}$ and $I_\mathrm{z}$ are the exoplanet and exozodi intensities, respectively.

System~(\ref{eq:I_system1}) can be reduced to the linear system
\begin{equation} \label{eq:I_system2}
\left \{
\begin{array}{l}
{(\widehat{\delta\psi_1})}^\ast \, (\widehat{P\phi}+\widehat{\psi}_0) + \widehat{\delta\psi_1} \, {(\widehat{P\phi}+\widehat{\psi}_0)}^\ast
= I_1 - I_0 - |\widehat{\delta\psi_1}|^2 \\
{(\widehat{\delta\psi_2})}^\ast \, (\widehat{P\phi}+\widehat{\psi}_0) + \widehat{\delta\psi_2} \, {(\widehat{P\phi}+\widehat{\psi}_0)}^\ast
= I_2 - I_0 - |\widehat{\delta\psi_2}|^2, \\
\end{array}
\right.
\end{equation}
where the exponent $\ast$ denotes the complex conjugate. Notice how the exoplanet and exozodi intensities have disappeared from the equations, demonstrating that faint objects do not affect the measurement process of stellar speckles. However, note that because of quantum noise, the planet detection can still be problematic if the exozodi is much brighter than the planet.

Now, system (\ref{eq:I_system2}) admits a unique solution if its determinant,
\begin{equation} \label{eq:delta}
\Delta \equiv {(\widehat{\delta\psi_1})}^\ast \, \widehat{\delta\psi_2} - \widehat{\delta\psi_1} \, {(\widehat{\delta\psi_2})}^\ast,
\end{equation}
is not zero, that is to say if
\begin{equation} \label{eq:I_condition}
|\widehat{\delta\psi_1}|\,|\widehat{\delta\psi_2}|\,\sin \! \left[ \arg(\widehat{\delta\psi_2}) - \arg(\widehat{\delta\psi_1}) \right] \neq 0.
\end{equation}

Condition (\ref{eq:I_condition}) tells us that the DM setting changes, $\delta\psi_1$ and $\delta\psi_2$, should modify the speckles differently in any given pixel, otherwise not enough information is secured to measure unambiguously $\widehat{P\phi}$ in this pixel. It should be expected for this method to work practically that the magnitude of the speckle modification be greater than the photon noise level.

We have not yet found a rigorous derivation of the optimum values for the amplitude $|\widehat{\delta\psi_1}|$ and $|\widehat{\delta\psi_2}|$, but a heuristic argument suggests to us that the optimum perturbations may be that $I_1 \approx I_0$ and $I_2 \approx I_0$. That is to say, the DM-induced speckle instensity pattern, taken by itself, should be approximately the same as the original speckle intensity pattern. Thus at each pixel we choose $|\widehat{\delta\psi_1}| \approx |\widehat{\delta\psi_2}| \approx \sqrt{I_0}$, with the caveat that neither should be zero to keep (\ref{eq:I_condition}) valid.

The phase of $\widehat{\delta\psi_1}$ does not matter, but the phase difference between $\widehat{\delta\psi_1}$ and $\widehat{\delta\psi_2}$ should be made as close to $\frac{\pi}{2}$ as possible to keep $\Delta$ from zero. Practically, this can be realized as follows:
\begin{enumerate}
\item Compute $\delta\psi_1$ stroke changes from (\ref{eq:coef2}) or (\ref{eq:energy6}) by replacing $\widehat{P\phi}$ by $\sqrt{I_0}\,e^{i\theta}$, where $\theta$ is a random phase;
\item Compute $\delta\psi_2$ stroke changes from (\ref{eq:coef2}) or (\ref{eq:energy6}) by replacing $\widehat{P\phi}$ by $\widehat{\delta\psi_1}\,e^{i\frac{\pi}{2}}$.
\end{enumerate}

Now that we have made sure that $\Delta \neq 0$, we derive finally
\begin{equation} \label{eq:I_solution}
\widehat{P\phi} = \frac{\widehat{\delta\psi_2} \, (I_1 - I_0 - |\widehat{\delta\psi_1}|^2) -
\widehat{\delta\psi_1} \, (I_2 - I_0 - |\widehat{\delta\psi_2}|^2)}{\Delta} - \widehat{\psi}_0.
\end{equation}

Equation (\ref{eq:I_solution}) shows that the initially unknown speckle field ($\widehat{P\phi}$) can be experimentally measured in just three exposures taken under identical circumstances but with different shapes imposed on the DM.

%
%
%
\section{Speckle nulling simulations} \label{sec:simulations}
%
%
\subsection{White speckle noise} \label{sub:sim_white}
In this section, we perform one- and two-dimensional simulations for the theoretical case of white speckle noise caused by phase aberrations only. The DM has 64 actuators and top-hat influence functions. Smoother influence functions have been tested and do not lead to qualitatively different results. A simulation with actual HCIT influence functions will be presented in the next section. The simulated portion of the pupil plane is made twice as big as the pupil by zero padding, so that every element of resolution in the image plane would be sampled by two detector pixels. This corresponds to the realistic case of a photon-starved exoplanet detection where read-out noise must be minimized.

%
%
\subsubsection{One-dimensional simulations}
\begin{figure}
\plotone{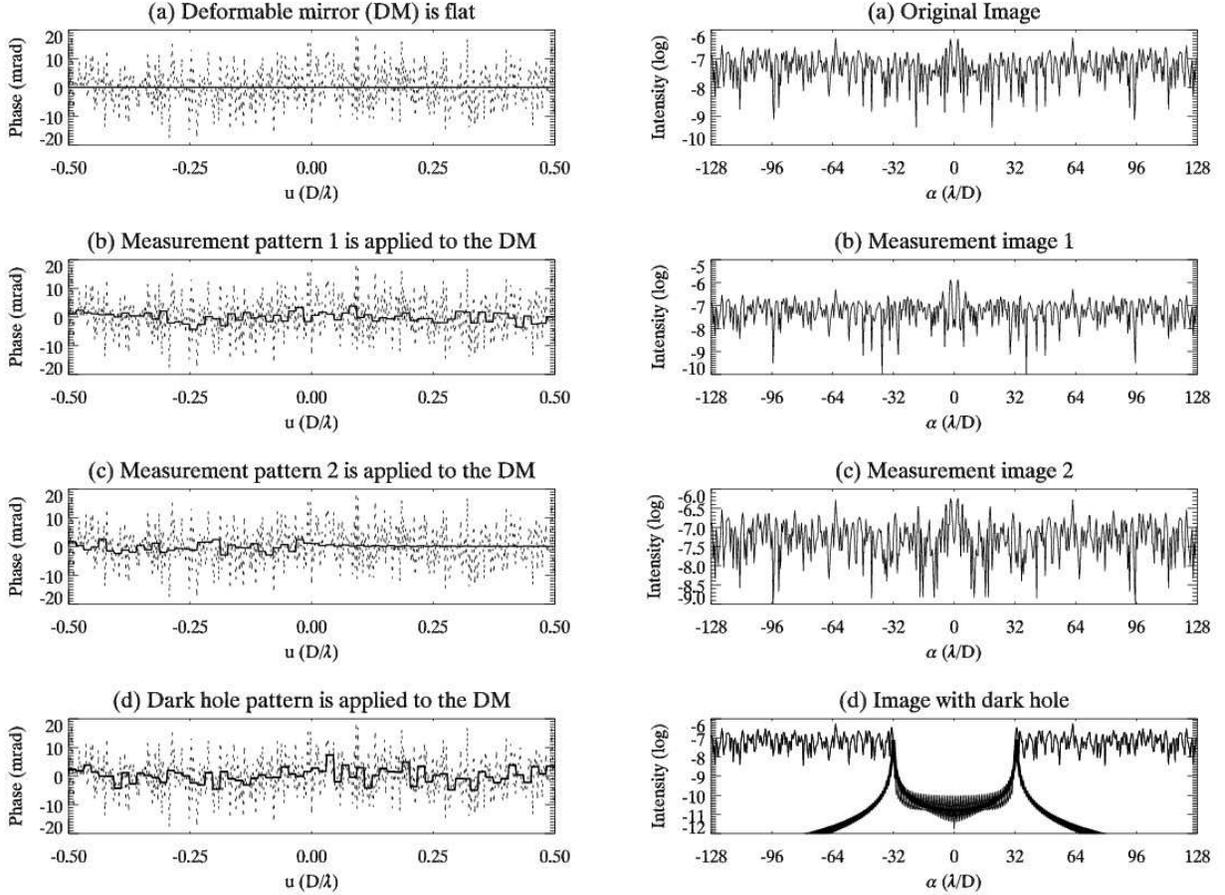}
\caption{Full one-dimensional speckle nulling simulation for a one-dimensional pupil with 64 actuators. Panels on the left show phase aberrations as dotted lines and their low spatial frequency approximations at the actuator scale as solid lines (negative of the patterns applied to the DM). Panels on the right show the corresponding images. The full algorithm is a four-step process: (a) original speckles are measured with the current DM shape (taken here to be flat); (b) and (c) the speckles are modified by driving the DM. At this point enough information has been gathered to deduce phase and amplitude aberrations of the wavefront; (d) low spatial frequency aberrations are corrected with the DM, canceling out speckles in the center of the image. Dark holes produced by the field nulling and energy minimization approaches (\S\ref{sub:field_nulling} \& \ref{sub:energy_min}) do not differ noticeably on the figure (thin solid line). A rough estimate based on the mean intensity prior to the dark hole creation (\S\ref{sub:sim_white}) is superimposed as a thick solid line.}
\label{fig:f1}
\end{figure}

Figure~\ref{fig:f1} shows a complete one-dimensional simulation including speckle field measurement (\S\ref{sub:measurement}) and speckle nulling with field nulling (\S\ref{sub:field_nulling}) and energy minimization (\S\ref{sub:energy_min}). The standard deviation of the phase aberrations is set to $\lambda/1000$. Intensities are scaled with respect to the maximum of the star PSF in the absence of a coronagraph. Ideal conditions are assumed: no photon noise, noiseless detector, and perfect precision in the control of DM actuators. Under these conditions, the speckle field is perfectly estimated, and the mean intensity in the DH is $5.8 \times 10^{-11}$ with field nulling and $1.4 \times 10^{-11}$ with energy minimization, i.e. about 1500 and 6500 times lower than the mean intensity outside the DH, respectively. Repeated simulations with different noise sequences show that energy minimization performs always better than field nulling by a factor of a few. Field nulling solved with SVD yields the same numerical solution as energy minimization (they differ by the last digit only), in agreement with the idea that they both minimize speckle energy.

%
%
\subsubsection{Dark hole depth estimate in one dimension}
In the one-dimensional case, it is easy to predict roughly the shape and the depth of the DH. The function $\widehat{P\phi} + \widehat{\psi}$ is band-limited since the pupil has a finite size. As the pupil linear dimension is $D/\lambda$, the maximum spatial frequency of $\widehat{P\phi} + \widehat{\psi}$ is $D/2\lambda$. Let us apply the sampling theorem at the Nyquist sampling frequency $D/\lambda$, and write
\begin{equation} \label{eq:1d1}
(\widehat{P\phi} + \widehat{\psi})(\alpha) = \sum_{n=-\infty}^{+\infty} \left[ \widehat{P\phi}_n + \widehat{\psi}_n \right] \mbox{sinc} \! \left( \frac{\alpha D}{\lambda} - n \right),
\end{equation}
where the subscript $n$ denotes the function value for $\alpha = n \frac{\lambda}{D}$.

Substituting $\alpha$ by $n \frac{\lambda}{D}$ and $d$ by $\frac{D}{N}$ leads to
\begin{equation}  \label{eq:1d2}
\widehat{P\phi}_n + \widehat{\psi}_n = \widehat{P\phi}_n + \hat{f}_n \sum_{k=0}^{N-1} a_k e^{-i \frac{2\pi k n }{N}}.
\end{equation}
The field nulling equation (\ref{eq:field1}) takes here the discrete form
\begin{equation} \label{eq:1d3}
\forall n \in \{ 0 \ldots N\!-\!1\}, \quad \widehat{P\phi}_n + \widehat{\psi}_n = 0
\quad \Longleftrightarrow \quad
a_k = \sum_{n=0}^{N-1} \left( -\frac{\widehat{P\phi}_n}{\hat{f}_n} \right) e^{i \frac{2\pi k n }{N}}.
\end{equation}
i.e. actuator strokes are computed thanks to an inverse FFT.

Let us now turn to the residual speckle field
\begin{equation} \label{eq:1d4}
(\widehat{P\phi} + \widehat{\psi})(\alpha) = \sum_{n=-\infty}^{-1} \widehat{P\phi}_n \: \mbox{sinc} \! \left( \frac{\alpha D}{\lambda} - n \right) + \sum_{n=N}^{+\infty} \widehat{P\phi}_n \: \mbox{sinc} \! \left( \frac{\alpha D}{\lambda} - n \right).
\end{equation}
Because the sinc function decreases rapidly with $\alpha$, the terms flanking the DH ($n=-1$ and $n=N$) should by themselves give the order of magnitude of the residual speckle field in the DH. In case of phase aberrations only and white noise, we have $|\widehat{P\phi}_{-1}|^2 \approx |\widehat{P\phi}_N|^2 \approx \overline{I_0}$, where $\overline{I_0}$ is the mean intensity in the image plane prior to the DH creation. Therefore, a crude estimate of the intensity profile in the DH should be
\begin{equation} \label{eq:1d5}
I_\mathrm{DH}(\alpha) \approx \overline{I_0} {\left[ \mbox{sinc} \! \left( \frac{\alpha D}{\lambda} +1 \right) + \mbox{sinc} \! \left( \frac{\alpha D}{\lambda} - N \right) \right]}^2.
\end{equation}

We have superimposed this approximation as a thick line in Fig.~\ref{fig:f1}. In this case the match is remarkable, but more simulations show that it is generally good within a factor of 10 only. Nevertheless, it demonstrates that the DH depth depends critically on the residual speckle field at its edges, hence on the decreasing rate of the complex aberration spectrum with spatial frequency. In that respect, a white spectrum is certainly the worst case. Equation (\ref{eq:1d5}) further indicates that the DH depth depends on the number of actuators: as $N$ is increased, the DH widens and gets deeper. With 8, 16, 32, and 64 actuators, (\ref{eq:1d5}) predicts $\overline{I_0}/\overline{I_\mathrm{DH}}$ to reach about 100, 300, 1000, and 4500. 

%
%
\subsubsection{Dark hole depth vs. search area}
\begin{figure}
\plotone{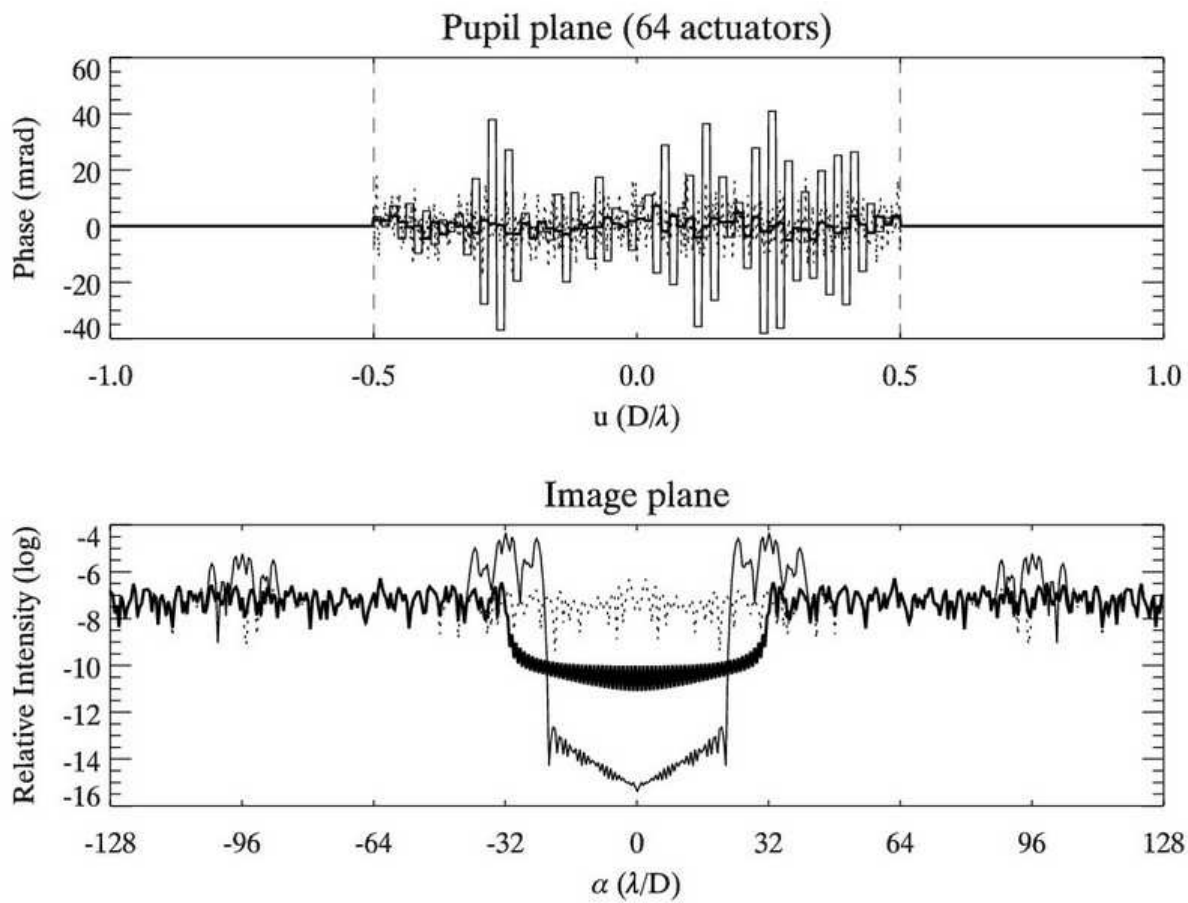}
\caption{The energy minimization algorithm makes it possible to push down the dark hole floor at the cost of some search area:  the original dark hole appears in thick solid line, while the deeper and narrower dark hole appears in thin solid line. In this particular example with phase aberrations only, the average floor is decreased from $1.4 \times 10^{-11}$ to $2.7 \times 10^{-15}$ (a factor of about 5200 in contrast) by reducing the dark hole size from $64\:\lambda/D$ to $44\:\lambda/D$ (31\,\% reduction). As explained in the text, this trade-off is obtained by increasing the amplitude of the highest spatial frequencies on the DM, making the dark hole's rim brighter at the same time.}
\label{fig:f2}
\end{figure}

As mentioned in \S\ref{sub:energy_min}, speckle nulling by energy minimization can be performed in a region narrower than the maximum DH. Figure~\ref{fig:f2} illustrates this point: by reducing the search area from 64 to 44 resels (31\,\% reduction), the DH floor was decreased from $1.4 \times 10^{-11}$ to $2.7 \times 10^{-15}$, i.e. a gain of about 5200 in contrast (further reducing the search area does not bring any significant gain). By giving up search space, one frees the degrees of freedom corresponding to the highest spatial frequency components on the DM pattern. These can be used to improve the DH depth at lower spatial frequency because of the PSF angular extension (this is essentially the same reason why high spatial frequency speckles limit the DH depth). As the search space is reduced, the leverage of these highest spatial frequency components decreases (PSF wings falling off). The energy minimization algorithm compensates by putting more energy at high frequency (see lower panel in Fig.~\ref{fig:f2}), which produces increasingly oscillatory DM patterns (see top panel in Fig.~\ref{fig:f2}) and increasingly brighter spots in the image (around $\pm 32 \frac{\lambda}{D}$ and $\pm 96 \frac{\lambda}{D}$ in bottom panel of Fig.~\ref{fig:f2}). Thus the trade-off range might be limited in practice by the maximum actuator stroke (currently $0.6\,\mu$m on the HCIT), and/or by the detector's dynamic range.

In two-dimensions, the trade-off limits are well illustrated by the following example: considering a $64\!\times\!64$ DM and a random wavefront, we find that the DH floor can be decreased from $2.4 \times 10^{-12}$ to $1.4 \times 10^{-13}$ (a factor of 17) if the search area is reduced from $64\!\times\!64$ to $60\!\times\!60$ resels (12\,\% reduction in area). This implies a maximum actuator stroke of 10\,nm and a detector dynamic range of $10^6$. A further reduction to $58\!\times\!58$ resels does not feature a lower DH floor ($2.1 \times 10^{-13}$), and would imply a maximum actuator stroke of $10\,\mu$m and a detector dynamic range of $10^{10}$. In this case, the leverage of the additionally freed high-spatial frequency components is so weak that the algorithm starts diverging.

%
%
\subsubsection{Two-dimensional simulations with phase and amplitude aberrations}
\begin{figure}
\plotone{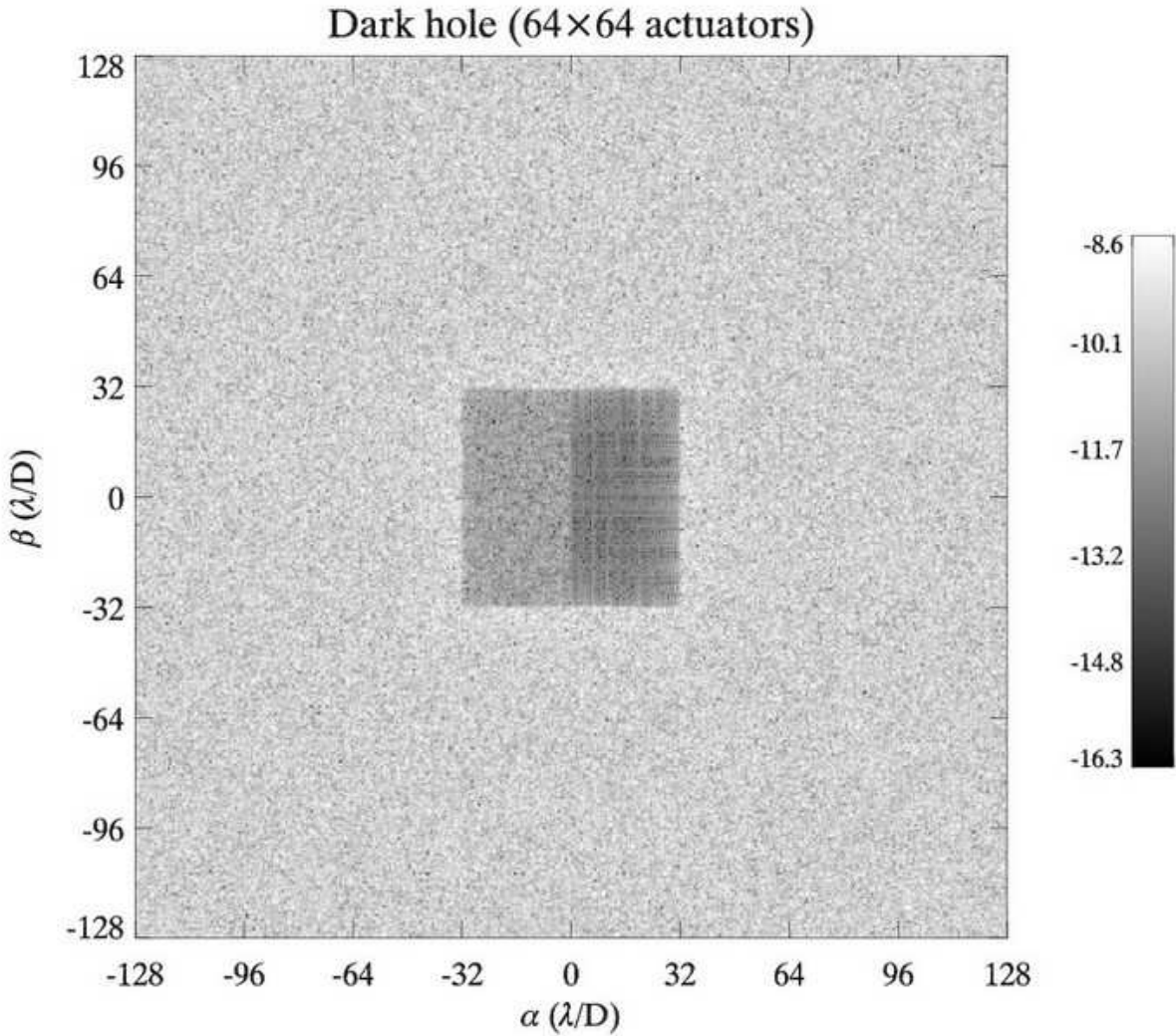}
\caption{Speckle nulling with the energy minimization algorithm (\S\ref{sub:energy_min}) for a square pupil in two dimensions. Grey levels code the logarithm of the intensity. Speckles are cleared from the central part of the image, creating a dark hole (or search area) suitable for the detection of faint companions. Phase aberrations are corrected in the full dark hole, while amplitude aberrations are corrected in the right part only and made worse by a factor of four in intensity in the left part. Thus the difference in intensity between the two sides of the dark hole gives a measure of the wavefront amplitude errors. In this simulation, the standard deviations of phase and amplitude aberrations are $\lambda/10^3$ and $\lambda/10^4$, respectively. The dark hole shape is a result of the actuator grid geometry on the DM, not of the pupil shape.}
\label{fig:f3}
\end{figure}

\begin{figure}
\plotone{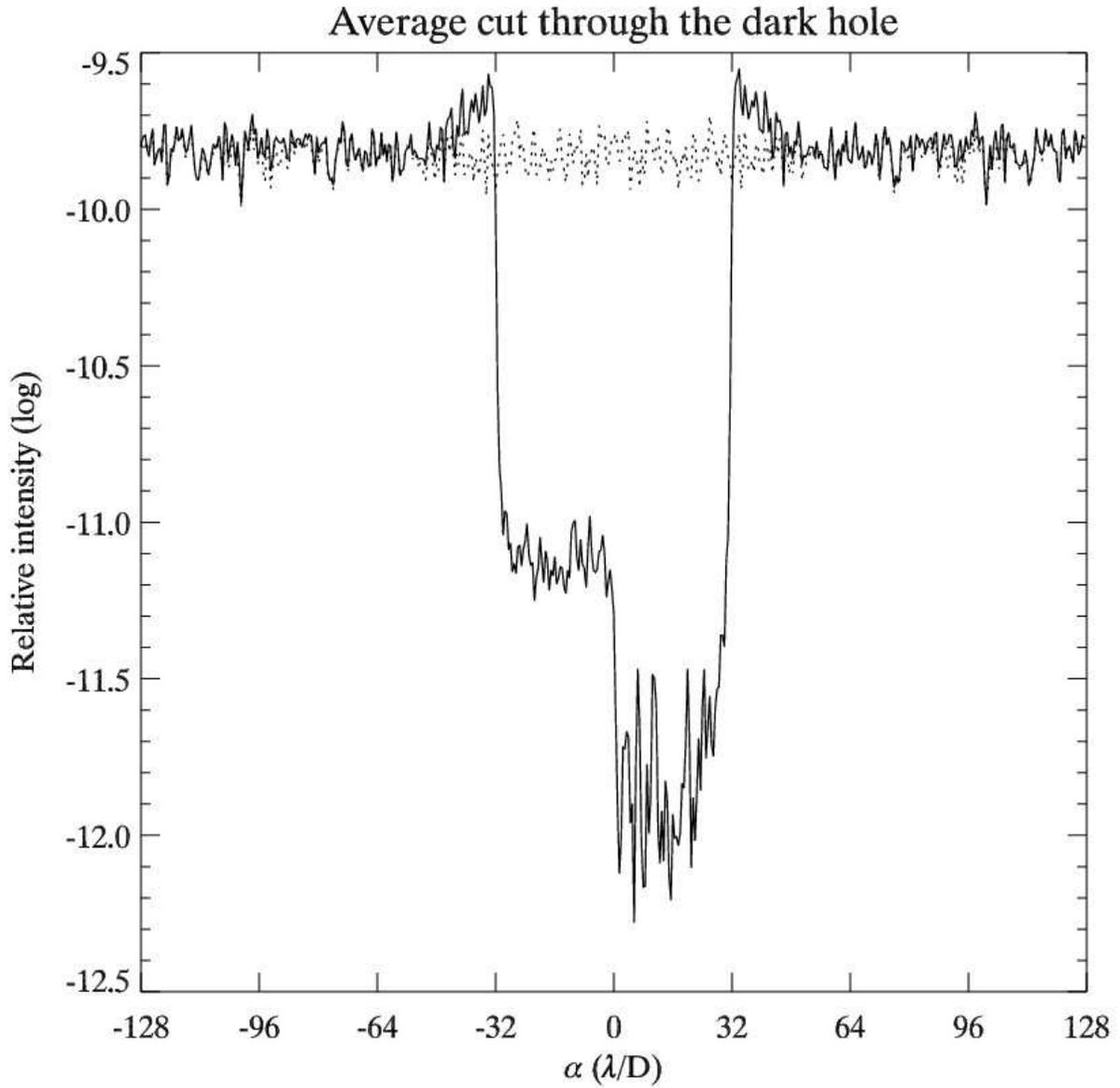}
\caption{Speckle nulling with the energy minimization algorithm (\S\ref{sub:energy_min}) in two dimensions. The solid curve is an average of Fig.~\ref{fig:f3} intensity over $\beta$. The dotted line represents the state prior to correction with the DM. Notice that the dark hole has a rim brighter than the background.}
\label{fig:f4}
\end{figure}

In Figs.~\ref{fig:f3}--\ref{fig:f4}, we show an example of two-dimensional speckle nulling with phase and amplitude aberrations for a square pupil. To reflect the fact that phase aberrations dominate amplitude aberrations in real experiments \cite[see][]{Trauger04}, the rms amplitude of amplitude aberrations is made ten times smaller than that of phase aberrations (the choice of a factor ten is arbitrary). The DH is split into two regions: in the right one ($\mathcal{H}^+$), amplitude and phase aberrations are corrected, whereas in the left one ($\mathcal{H}^-$), phase aberrations are corrected and amplitude aberrations are made worse by a factor of four in intensity.

%
%
\subsection{Realistic speckle noise} \label{sub:sim_real}
%
%
\subsubsection{Power spectral density of phase aberrations}
\begin{figure}
\plotone{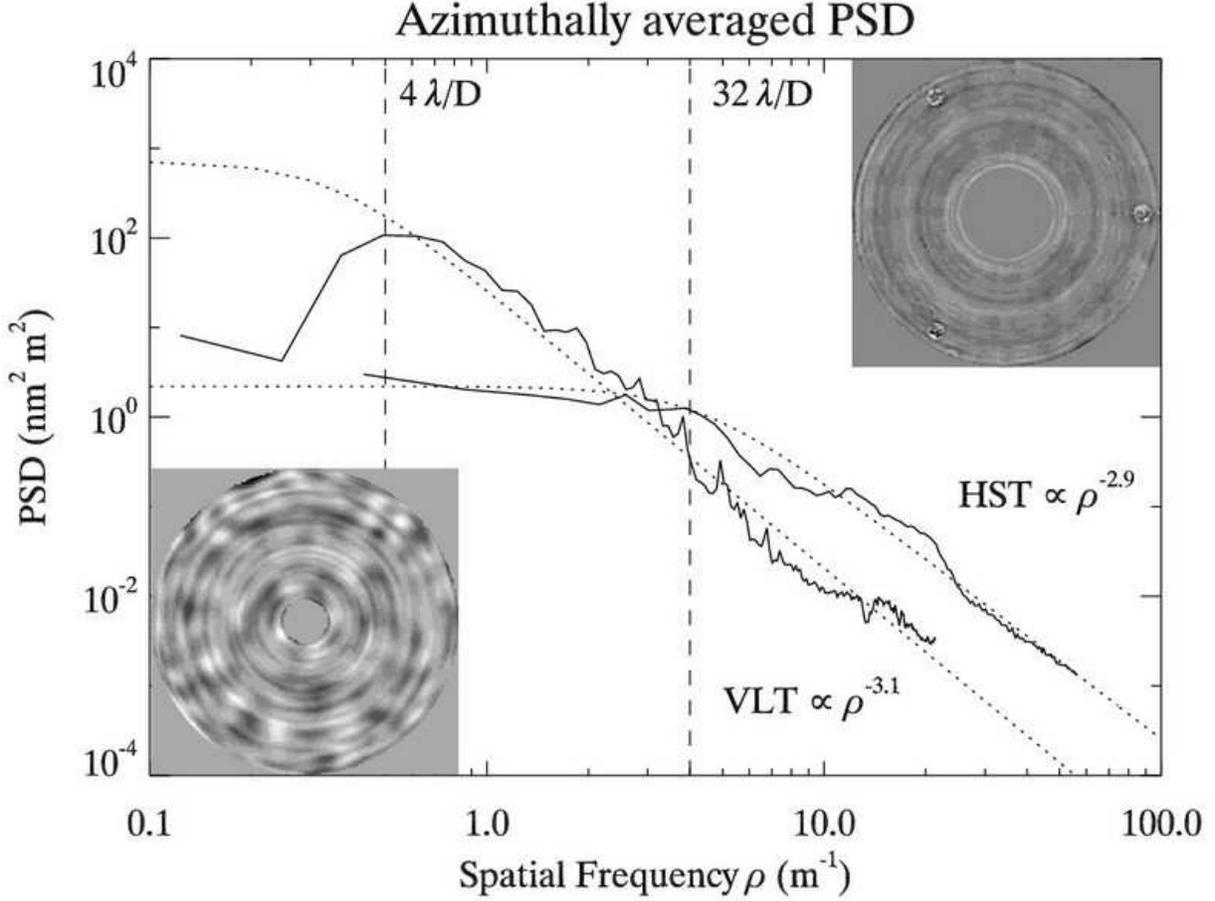}
\caption{Azimuthally averaged power spectral density (PSD) of phase aberrations for (i) the 8.2-m primary mirror of a VLT unit telescope (Antu), and (ii) the combination of the primary and secondary mirrors of Hubble. Both data sets appear in solid line, and are fitted with model (\ref{eq:psd}) drawn in dotted line. Model parameters are listed in Table~\ref{tab:t1}. The dashed lines indicate the boundaries for spatial frequencies leading to speckles in the 4--32 $\lambda/D$ region of the image plane for an 8-m mirror. The lower-left and upper-right insets are the VLT phase map (courtesy of ESO) and HST phase map \citep{Krist95}, respectively.}
\label{fig:f5}
\end{figure}

With the $3.5\!\times\!8$-m TPF-C primary mirror in mind, we have studied the phase aberration map of an actual 8-m mirror: the primary mirror of Antu, the first 8.2-m unit telescope of ESO's Very Large Telescope (VLT). This phase map\footnote{It can be found by courtesy of ESO at http://www.eso.org/projects/vlt/unit-tel/m1unit.html.} was obtained with the active optics system on, and is characteristic of zonal errors (aberrations which cannot be fitted by low-order Zernike-type polynomials). It can be seen in Fig.~\ref{fig:f5} that the azimuthally averaged power spectral density (PSD) of such a map is well represented by
\begin{equation} \label{eq:psd}
\mbox{PSD}(\rho) = \frac{\mbox{PSD}_0}{1+{(\rho/\rho_c)}^x},
\end{equation}
where $\rho = \sqrt{\alpha^2+\beta^2}$. Values for PSD$_0$, $\rho_c$ and $x$ are listed in Table~\ref{tab:t1}. For comparison, the same treatment has been applied to the Hubble Space Telescope (HST) zonal error map from \citet{Krist95}.

We conclude from this study that \emph{a realistic phase aberration PSD for an 8-m mirror decreases as the third power of the spatial frequency}. The standard deviation of the VLT phase map is 20.9~nm (18.5~nm for HST). The square root of the power of phase aberrations in the 0.5--4~m$^{-1}$ spatial frequency range (4--32~$\lambda/D$ for an 8-m mirror) is 19.4~nm, i.e. about $\lambda/25$ at 500~nm, clearly not in the validity domain of our linear approximation.

%
%
\subsubsection{One-dimensional simulation}
\begin{figure}
\plotone{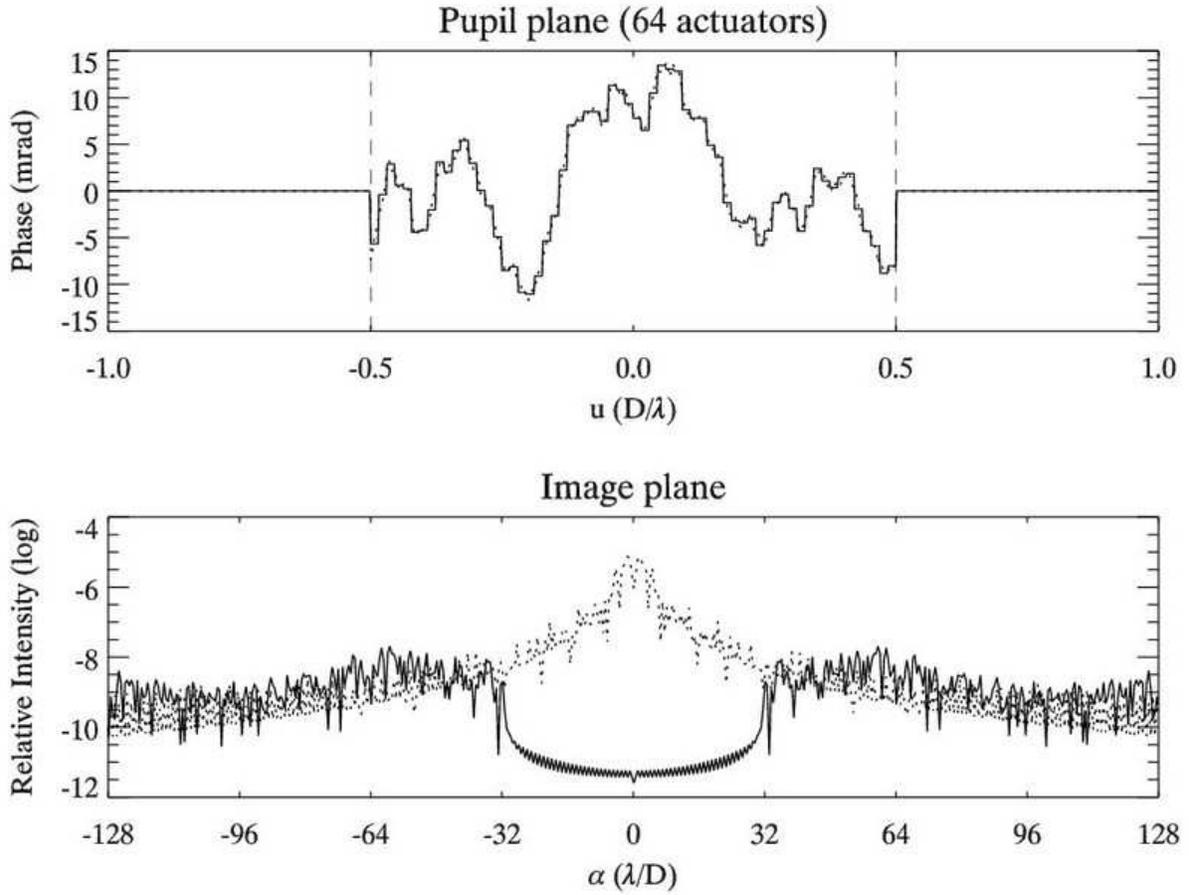}
\caption{One-dimensional speckle nulling as computed by energy minimization (\S\ref{sub:energy_min}) with a phase aberration PSD given by (\ref{eq:psd}). Values for $x$ and $\rho_c$ are the same as for the VLT; PSD$_0$ is such that the standard deviation of phase aberrations is the same as in Fig.~\ref{fig:f1} ($\lambda/1000$). These realistic phase aberrations lead to deeper dark holes than in the hypothetical white-noise case of Fig.~\ref{fig:f1}d.}
\label{fig:f6}
\end{figure}

Figure~\ref{fig:f6} shows a simulation performed in the same conditions as Fig.~\ref{fig:f1}, but with a VLT-like PSD. The PSD is scaled so that the standard deviation of phase aberrations is equal to $\lambda/1000$. The average DH floor is now $5.3 \times 10^{-12}$, six orders of magnitude below the intensity peak in the original image! In agreement with \S\ref{sub:sim_white}, we find that \emph{the DH's depth depends critically on the magnitude of the speckle field at the edge of the DH, hence on the decrease of the phase aberration PSD with spatial frequency}.

%
%
\subsubsection{Two-dimensional simulation}
\begin{figure}
\plotone{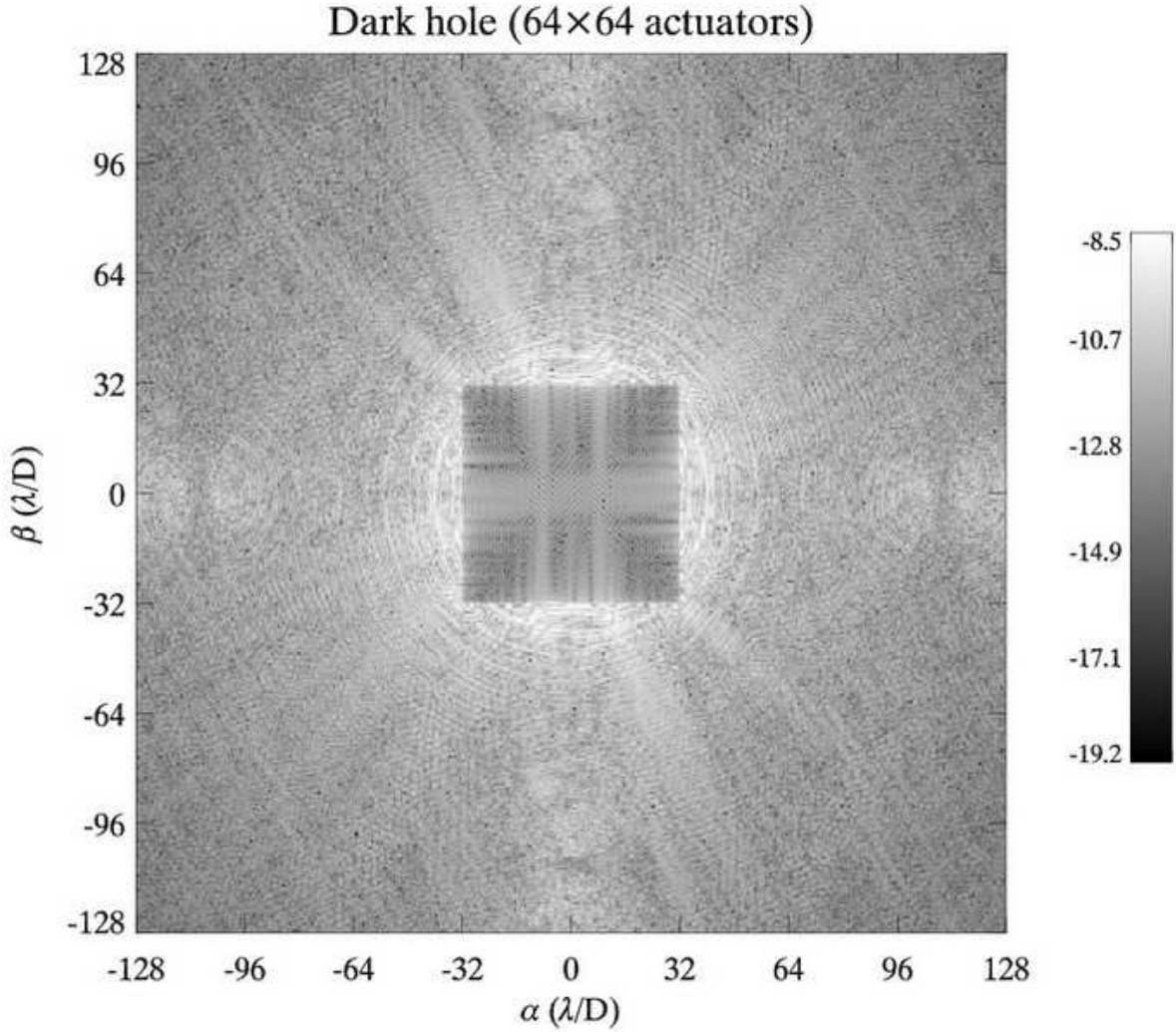}
\caption{Speckle nulling with the field nulling algorithm (\S\ref{sub:field_nulling}) with a VLT-like PSD instead of white noise. The standard deviation of phase aberrations is fixed to $\lambda/1000$. Actual HCIT influence functions are used. The average DH floor is $5.9 \times 10^{-12}$.}
\label{fig:f7}
\end{figure}

\begin{figure}
\plotone{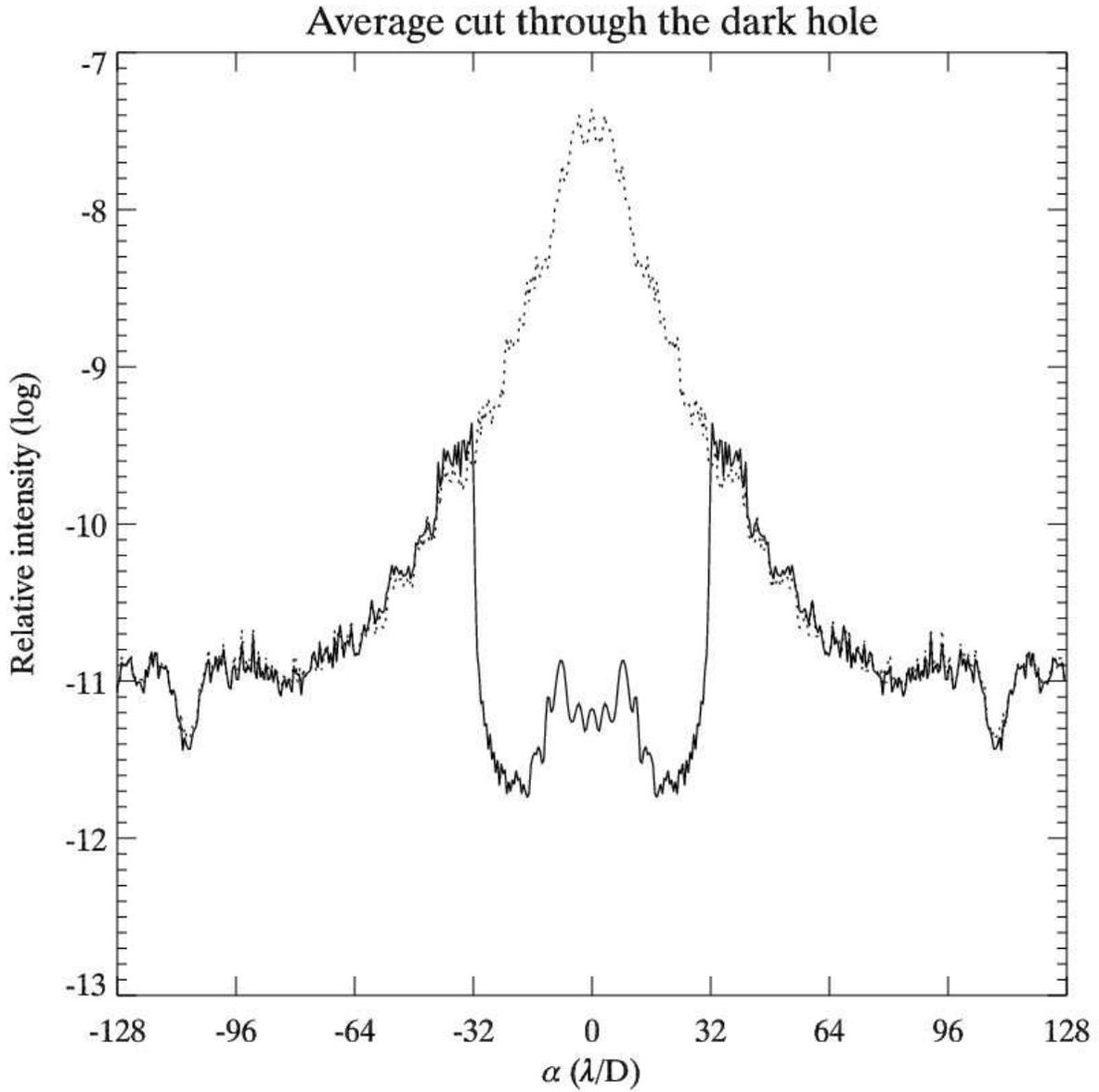}
\caption{Speckle nulling with the field nulling algorithm (\S\ref{sub:field_nulling}) with a VLT-like PSD instead of white noise. The solid curve is an average of Fig.~\ref{fig:f7} intensity over $\beta$. The dotted line represents the state prior to correction with the DM.}
\label{fig:f8}
\end{figure}

For the two-dimensional simulation in Figs.~\ref{fig:f7}--\ref{fig:f8}, we have kept the original VLT phase map and circular pupil, but scaled the standard deviation of phase aberrations to $\lambda/1000$. In addition, we have used the actual HCIT influence functions from \citet{Trauger03}. The average DH floor is then $5.9 \times 10^{-12}$ with field nulling (case shown), and $7.1 \times 10^{-11}$ with energy minimization. The worse performance of the second method reflects the cost of the variable separation hypothesis, only accurate to within 5\,\% for the HCIT. Note that the DH retains its square shape with a circular pupil, as the DH shape is fixed by the actuator grid geometry on the DM (a square grid of constant spacing in our case).

%
%
\section{Discussion} \label{sec:discussion}
%
%
\subsection{Quantum and read-out noise} \label{sub:discuss_noise}
In \S\ref{sec:simulations}, we presented noise-free simulations. To give an idea of the effect of quantum and read-out noises, let us consider a sun-like star at 10~pc observed by a $3.5\!\times\!8$~m space telescope with a 5\,\% overall efficiency. In a 100~nm bandwith centered at 600~nm, the telescope collects about $2 \times 10^{12}$ photo-electrons in one-hour exposures. Considering the quantum noise, a 1~e$^-$ read-noise and ignoring chromatic effects, simulations of sequences of four one-hour exposures show that the average DH floor in Fig.~\ref{fig:f1} would jump from $1.4 \times 10^{-11}$ to $2.7 \times 10^{-10}$, whereas the average DH floor in Fig.~\ref{fig:f6} would jump from $5.2 \times 10^{-12}$ to $3.2 \times 10^{-11}$. 

%
%
\subsection{Validity of the linear approximation} \label{sub:discuss_linearity}
In practice, our speckle nulling process will work as stated provided Eq.~(\ref{eq:E_im}) holds, that is to say if ${|P\phi+\psi| \gg \frac{1}{2}|P\phi^2|}$. If $c$ is the improvement in contrast with respect to the speckle floor and $\sigma_\phi$ the standard deviation of wavefront aberrations in radians, this condition translates into ${\sigma_\phi/\sqrt{c} \gg \sigma_\phi^2/\sqrt{2}}$, or ${\sigma_\phi \ll \sqrt{2/c}}$. In terms of optical path difference, the standard deviation should then be much less than $\lambda/(\pi \sqrt{2c}) = \lambda/140$ for $c = 10^3$. This is why we considered $\lambda/1000$ rms wavefronts in our simulations. As the wavefront will probably not be of this quality at the start, the speckle nulling method presented here is intended to be used in the course of observations, after a first phase where the bulk of the aberrations have been taken out.

When the linear approximation breaks down, three images with different DM settings still provide enough information about the aberrations, so that a DH could be created thanks to a global non-linear analysis of these images \citep{Borde04}. \cite{Malbet95} also explored non-linear solutions, but with many more iterations ($\approx 20$).

%
%
\subsection{Real coronagraphs}  \label{sub:discuss_coronagraphs}
Dwelling on the validity of Eq.~(\ref{eq:E_im}), real coronagraphs would not only remove the direct image of the star ($\widehat{P}$), they would also modify the speckle field ($\widehat{P\phi}$) and the DM phase function ($\widehat{P\psi}$). This can be easily incorporated in the theory. A more delicate point is that real coronagraphs are not translation-invariant systems. As a consequence, effective influence functions as seen from behind the coronagraph will vary over the pupil. For image-plane coronagraphs with band-limited sinc masks \cite[][\S4]{Kuchner02}, we estimate this variation to be of the order of 10\,\%, assuming $\epsilon = 0.1$ and 64 actuators. Only energy minimization, not field nulling (unless solved with SVD), can accomodate this effect.

%
%
\subsection{Actuator stroke precision}  \label{sub:discuss_actuators}
What about the precision at which actuators should be controlled? As a consequence of the linearity of (\ref{eq:energy3}), the DH depth depends quadratically on the precision on the actuator strokes. We deduce -- and this is confirmed by simulations -- that a four orders of magnitude deep DH can only be obtained if the strokes are controlled at a 1\,\% precision, i.e. 6\,pm rms with $\lambda/1000$ aberrations at 600\,nm. This precision corresponds to the current resolution of the actuator drivers on the HCIT.

%
%
\subsection{Instrumental stability}  \label{sub:discuss_stability}
Regarding instrumental stability, we assumed that the instrument would remain perfectly stable during the four-step process. However, despite the foreseen thermal and mechanical controls of the spacecraft, very slow drifts during the few hours of single exposures should be expected. Therefore we intend to study in a subsequent paper how to incorporate a model of the drifts in our method. The exact parameters of this model would be derived from a learning phase after the launch of the spacecraft.

%
%
\subsection{Chromaticity}  \label{sub:discuss_chromaticity}
We have not considered the effect of chromaticity. Let us point out that phase aberrations due to mirror surface errors scale with wavelength, so the correction derived from one wavelength would apply to all wavelengths. This is unfortunately not the case for amplitude aberrations. Although these are weaker than phase aberrations, a degradation of the correction should be expected in polychromatic light. Moreover, polychromatic wavefront sensing will require a revised theory as speckles will move out radially in proportion to the wavelength.

%
%
%
\section{Conclusion and future work} \label{sec:conclusion}
In this paper, we presented two techniques to optimally null out speckles in the central field of an image behind an ideal coronagraph in space. The measurement phase necessitates only three images, the fourth image being fully corrected. Depending on the number of actuators and the desired search area, the gain in contrast can reach several orders of magnitude.

These techniques are intended to work in a low aberration regime, such as in the course of observations after an initial correction phase. They are primarily meant to be used in space but could be implemented in a second-stage AO system on ground-based telescopes. Out of these two methods, the speckle energy minimization approach seems to be the more powerful and flexible: (i) it offers the possibility to trade off some search area against an improved contrast, and (ii) it can accomodate influence function variations over the pupil (necessary with real coronagraphs). If influence functions feature the required symmetry (variable separation), it is computationally very effective, but is otherwise still better than SVD. 

Since the principles underlying these speckle nulling techniques are general, it should be possible to use them in conjunction with most coronagraph designs, including those with band-limited masks \citep{Kuchner02}, pupil-mapping \citep{Guyon05a,Vanderbei05}, and shaped pupils \citep{Kasdin03}. It is our intent to complete our work by integrating in our simulations models of these coronagraphs, and to carry out experiments with the HCIT.

In addition, we will seek to incorporate in the measurement theory a linear model for the evolution of aberrations, and we will work toward a theory accommodating the spectral bandwidth needed for the detection and spectroscopy of terrestrial planets.

%
%
\acknowledgments

We wish to thank the anonymous referee for his insightful comments that helped to improve greatly the content of this paper. We acknowledge many helpful discussions with Chris Burrows, John Trauger, Joe Green, and Stuart Shaklan.  This work was performed in part under contract 1256791 with the Jet Propulsion Laboratory (JPL), funded by NASA through the Michelson Fellowship Program, and in part under contract 1260535 from JPL. JPL is managed for NASA by the California Institute of Technology. This research has made use of NASA's Astrophysics Data System.

%
%
\appendix
%
%
\section{Formulation of energy minimization in the pupil plane} \label{app:global}
In this appendix, we show that energy minimization can be formulated in the pupil plane as well. Note that no measurement takes place in the pupil plane: the aberration function $P\phi$ is obtained by the inverse FT of $\widehat{P\phi}$ which is still measured as described in \S\ref{sub:measurement}. Although we present here a solution for phase aberrations, amplitude aberrations can be corrected in half of the domain without changing the formalism, exactly as explained in \S\ref{sub:energy_min}.

By replacing $\widehat{E}$ with its expression as a FT, the dark hole energy reads
\begin{equation} \label{eq:app1}
\mathcal{E} = \int\!\!\!\int_{\mathcal{H}} \left (
\int\!\!\!\int_{\mathcal{P}} E(u,v) \, e^{-i2\pi(u\alpha+v\beta)} \mathrm{d}u \, \mathrm{d}v
\int\!\!\!\int_{\mathcal{P}} E^\ast(u',v') \, e^{i2\pi(u'\alpha+v'\beta)} \mathrm{d}u' \mathrm{d}v'
\right ) \mathrm{d}\alpha \, \mathrm{d}\beta.
\end{equation}

Now we invert the integration order and integrate over $\mathcal{H}$ to get
\begin{equation} \label{eq:app2}
\mathcal{E} = \frac{1}{4d^2} \int\!\!\!\int_{\mathcal{P}} E(u,v) \int\!\!\!\int_{\mathcal{P}} E^\ast(u',v') \, h(u'-u)\, h(v'-v) \,
\mathrm{d}u \, \mathrm{d}v \, \mathrm{d}u' \mathrm{d}v',
\end{equation}
where $h(x-y) \equiv  \, \mbox{sinc} [\lambda(x-y)/2d]$. The energy is minimized when
\begin{equation} \label{eq:app3}
\forall (k,l), \:
\int\!\!\!\int_{\mathcal{P}} f_{kl}(u,v) \int\!\!\!\int_{\mathcal{P}} \Big[ \phi(u',v') + \psi(u',v') \Big] \, h(u'-u)\, h(v'-v) \,
\mathrm{d}u \, \mathrm{d}v \, \mathrm{d}u' \mathrm{d}v' = 0.
\end{equation}

In the next steps, we first replace $\psi$ with (\ref{eq:psi}), then $f_{kl}(u,v)$ with its tensor product $g_k(u) \, g_l(v)$ in order to integrate separately the variables $(u, u')$ and $(v, v')$. The final result reads
\begin{eqnarray} \label{eq:app4}
\forall (k,l) \in {\{0 \ldots N\!-\!1\}}^2, & & \sum_{n=0}^{N-1} G_{kn} \sum_{m=0}^{N-1} a_{nm} \, G_{ml} = \Phi_{kl} \\
\quad \mbox{where} \quad
G_{ij} & = & \int\!\!\!\int_{\mathcal{P}} g_i(x) \, g_j(y) \, h(x-y) \, \mathrm{d}x \, \mathrm{d}y \nonumber \\
\quad \mbox{and} \quad
\Phi_{kl} & = & \int\!\!\!\int_{\mathcal{P}} g_k(u) \, g_l(v) \int\!\!\!\int_{\mathcal{P}} \phi(u',v') \, h(u'-u)\, h(v'-v) \, \mathrm{d}u \, \mathrm{d}v \, \mathrm{d}u' \mathrm{d}v'. \nonumber
\end{eqnarray}
System (\ref{eq:app4}) has a form identical to (\ref{eq:energy5}) and can be solved with the same technique.

%
%

%
%
\clearpage
%
%

%
%

%
%

%
%

%
%

%
%

%
%

%
%

%
%
\clearpage
%
%
\begin{table}
\begin{center}
\caption{Parameter values for the azimuthally averaged PSD model of VLT and HST phase maps. \label{tab:t1}}
\begin{tabular}{cccc}
\tableline\tableline
    & PSD$_0$ (nm$^2$~m$^2$) & $\rho_c$ (m$^{-1}$) & $x$ \\
\tableline
HST &  2.2                   & 4.3                         & 2.9      \\
VLT &  720                   & 0.35                        & 3.1      \\
\tableline
\end{tabular}
\end{center}
\end{table}


\begin{thebibliography}{}

\bibitem[Angel(2003)]{Angel03}
Angel, R.  2003, ESA SP, 539, 221

\bibitem[Bord\'e, Traub, \& Trauger(2004)]{Borde04}
Bord\'e, P., Traub, W., \& Trauger, J.  2004, Proceedings of the Second TPF/Darwin conference, meeting held in San Diego, USA, July 26--29, 2004, http://planetquest1.jpl.nasa.gov/TPFDarwinConf/confProceedings.cfm

\bibitem[Charbonneau et al.(2005)]{Charbonneau05}
Charbonneau, D., Allen, L., Megeath, S., Torres, G., Alonso, R., Brown, T., Gilliland, R., Latham, D., Mandushev, G., O'Donovan, F., \& Sozzetti, A. 2005, \apj, 626, 523

\bibitem[Chauvin et al.(2004)]{Chauvin04}
Chauvin, G., Lagrange, A.-M., Dumas, C., Zuckerman, B., Mouillet, D., Song, I., Beuzit, J.-L., \& Lowrance, P.  2004, \aap, 425, L29

\bibitem[Chauvin et al.(2005)]{Chauvin05}
Chauvin, G., Lagrange, A.-M., Dumas, C., Zuckerman, B., Mouillet, D., Song, I., Beuzit, J.-L., \& Lowrance, P.  2005, \aap, 438, L25

\bibitem[Chelli(2005)]{Chelli05}
Chelli, A.  2005, \aap, 441, 1205

\bibitem[Coulter(2004)]{Coulter04}
Coulter, D.  2004, \procspie, 5487, 1207

\bibitem[Des Marais et al.(2002)]{DesMarais02}
Des Marais, D., Harwit, M., Jucks, K., Kasting, J., Lin, D., Lunine, J., Schneider, J., Seager, S., Traub, W., \& Woolf, N.  2002, Astrobiology, 2, 153

\bibitem[Deming et al.(2005)]{Deming05}
Deming, D., Seager, S., Richardson, L., \& Harrington, J.  2005, \nat, 434, 740

\bibitem[Green et al.(2003)]{Green03}
Green, J., Basinger, S., Cohen, D., Niessner, A., Redding, D., Shaklan, S., \& Trauger, J.  2003, \procspie, 5170, 38

\bibitem[Guyon et al.(2005)]{Guyon05a}
Guyon, O., Pluzhnik, E., Galicher, R., Martinache, F., Ridgway, S., \& Woodruff, R.  2005, \apj, 622, 744

\bibitem[Guyon(2005)]{Guyon05b}
Guyon, O.  2005, \apj, 629, 592

\bibitem[Hsu(1967)]{Hsu67}
Hsu, H.  1967, Outline of Fourier Analysis, Unitech Division

\bibitem[Karlsson et al.(2004)]{Karlsson04}
Karlsson, A., Wallner, O., Perdigues Armengol, J., \& Absil, O.  2004, \procspie, 5491, 831

\bibitem[Kasdin et al.(2003)]{Kasdin03}
Kasdin, N., Vanderbei, R., Spergel, D., \& Littman, M.  2003, \apj, 582, 1147

\bibitem[Krist \& Burrows(1995)]{Krist95}
Krist, J., \& Burrows, C.  1995, \ao, 34, 4951 

\bibitem[Kuchner \& Traub(2002)]{Kuchner02}
Kuchner, M., \& Traub, W.  2002, \apj, 570, 900

\bibitem[Labeyrie(1995)]{Labeyrie95}
Labeyrie, A.  1995, \aap, 298, 544

\bibitem[L\"ofdahl \& Scharmer(1994)]{Lofdahl94}
L\"ofdahl, M., \& Scharmer, G.  1994, \aaps, 107, 243

\bibitem[Malbet, Yu, \& Shao(1995)]{Malbet95}
Malbet, F., Yu, J., \& Shao, M.  1995, \pasp, 107, 386

\bibitem[Neuh\"{a}user et al.(2005)]{Neuhauser05}
Neuh\"{a}user, R., Guenther, E., Wuchterl, G., Mugrauer, M., Bedalov, A., \& Hauschildt, P.  2005, \aap, 435, L13

\bibitem[Press et al.(2002)]{Press02}
Press, W., Teukolsky, S., Vetterling, W., \& Flannery, B.  2002, Numerical recipes in C: the art of scientific computing, Cambridge University Press

\bibitem[Quirrenbach(2005)]{Quirrenbach05}
Quirrenbach, A. 2005, Conclusions from a workshop held February 02-06, 2004 at Leiden University, astro-ph/0502254

\bibitem[Trauger et al.(2003)]{Trauger03}
Trauger, J., Moody, D., Gordon, B., G\"ursel, Y., Ealey, M., \& Bagwell, R.  2003, \procspie, 4854, 1

\bibitem[Trauger et al.(2004)]{Trauger04}
Trauger, J., Burrows, C., Gordon, B., Green, J., Lowman, A., Moody, D., Niessner, A., Shi, F., \& Wilson, D.  2004, \procspie, 5487, 1330

\bibitem[Vanderbei \& Traub(2005)]{Vanderbei05}
Vanderbei, R., \& Traub, W.  2005, \apj, 626, 1079

\end{thebibliography}
\end{document}